\documentclass[entropy,article,accept,moreauthors,pdftex,10pt,a4paper]{mdpi} 
\usepackage{color}
\usepackage{soul}

\firstpage{1} 
\articlenumber{x}
\doinum{10.3390/------}
\pubvolume{xx}
\pubyear{2021}
\copyrightyear{2021}
\externaleditor{Academic Editor: name}
\history{Received: date; Accepted: date; Published: date}

\newcommand{\e}{\mathrm{e}}

\usepackage{glossaries}

\Title{Multi-particle interference in an electronic Mach-Zehnder interferometer}

\Author{Janne Kotilahti,$^{1}$ Pablo Burset,$^{1,2}$ Michael Moskalets,$^{3}$ and Christian Flindt$^{1}$}

\AuthorNames{Janne Kotilahti, Pablo Burset, Michael Moskalets and Christian Flindt}

\address{%
$^{1}$ Department of Applied Physics, Aalto University, Finland\\
$^{2}$ Department of Theoretical Condensed Matter Physics, Universidad Aut\'onoma de Madrid, Madrid, Spain\\
$^{3}$ Department of Metal and Semiconductor Physics, NTU ``Kharkiv Polytechnic Institute'', Kharkiv, Ukraine}

\abstract{The recent development of dynamic single-electron sources makes it possible to observe and manipulate the quantum properties of individual charge carriers in mesoscopic circuits. Here, we investigate multi-particle effects in an electronic Mach-Zehnder interferometer driven by dynamic voltage pulses. To this end, we employ a Floquet scattering formalism to evaluate the interference current and the visibility in the outputs of the interferometer. An injected multi-particle state can be described by its first-order correlation function, which we decompose into a sum of elementary correlation functions that each represent a single particle. Each particle in the pulse contributes independently to the interference current, while the visibility (determined by the maximal interference current) exhibits a Fraunhofer-like diffraction pattern caused by the multi-particle interference between different particles in the pulse.  For a sequence of multi-particle pulses, the visibility resembles the diffraction pattern from  a grid, with the role of the grid and the spacing between the slits being played by the pulses and the time delay between them. Our findings may be observed in future experiments by injecting multi-particle pulses into an electronic Mach-Zehnder interferometer.
}


\begin{document}

\section{Introduction}

Quantum-coherent circuits based on mesoscopic conductors~\cite{Bocquillon_2013b} combined with dynamic single-electron emitters~\cite{Splettstoesser_2017,Bauerle_2018b} have paved the way for experiments on high-frequency quantum transport~\cite{Gabelli_2006,Feve_2007,Bocquillon_2013,Dubois_2013,Jullien_2014,Roussely_2018} and are holding great promises for future quantum technologies. Advances in nanotechnology have made it possible to fabricate highly pure samples and cool them to sub-Kelvin temperatures, where the phase coherence of the charge carriers is preserved over large enough distances to exploit and control their quantum behavior. By utilizing the quantum Hall effect in a strong magnetic field, the electrons can be forced to move along chiral edge states, as if traveling on rail tracks~\cite{QSHI_2007,Brune_2010,Sullivan_2011,Brune_2012}. In addition, several dynamic single-electron emitters have been developed. Single electrons can now be injected into a quantum-coherent circuit using driven mesoscopic capacitors~\cite{Buttiker_1993,Feve_2007,Bocquillon_2013}, dynamic quantum dots~\cite{Blumenthal_2007,Kataoka_2013,Fletcher_2019} or by applying Lorentzian-shaped pulses to a contact~\cite{Dubois_2013,Jullien_2014} as first envisioned by Levitov and co-workers~\cite{Levitov_1996,Levitov_1997,Keeling_2006}. When operated in the giga-Hertz regime, these setups make it possible to investigate and observe the quantum behavior of a single or a few electrons in a circuit.

Interference of particles is one central prediction of quantum mechanics, which distinguishes it from classical physics. Interference patterns can develop because the states of quantum particles are described by wave functions. The quantum nature of photons was clarified in pioneering work such as the Hanbury Brown-Twiss~\cite{HanburyBrown_1956,Brown_1956} and Hong-Ou-Mandel~\cite{Hong_1987} interference experiments and by  the quantum mechanical theory of light by Glauber based on correlation functions \cite{Glauber1963photon,Glauber1963coherence,Glauber1963quantum}. This understanding led to the development of quantum optics, which uses beam splitters, mirrors, and other optical components to investigate the statistics and coherence of photons, and eventually paved the way for quantum technologies such as quantum communication with photons~\cite{Ursin_2007}. 

Quantum optics-like experiments can now be conducted with electrons, which has led to the field called electron quantum optics~\cite{Bocquillon_2012,Bocquillon_2013b,Glattli_2018}. Electrons in a circuit can be manipulated using electronic analogues of various optical components. For example, quantum point contacts~\cite{Foxon_1988,Jones_1988} may function as beam splitters and  serve as the elementary building blocks of electronic Mach-Zehnder~\cite{Henny_1999,Oliver_1999,Oberholzer_2000,Ji_2003,Neder_2007,Neder_2007b,Mailly_2008,Strunk_2008,West_2009,Tewari_2016,Roche_2012,Roulleau_2021} and Fabry-P\'{e}rot interferometers~\cite{Liang_2001,Kretinin_2010,Gaury_2014}. Obviously, the differences between electrons and photons have clear implications for such experiments. For instance, due to the Pauli exclusion principle, two electrons cannot occupy the same single-particle state unlike photons. Furthermore, electrons in a mesoscopic conductor propagate on top of the underlying Fermi sea in contrast to photons that typically travel in vacuum. Also, in some situations electrons interact strongly with each other via the mutual Coulomb interactions~\cite{Grenier_2013,Ferraro_2014,Ferraro_2015,Rech_2017,Ferraro_2017,Ferraro_2017b,Cabart_2018,Ronetti_2018,Rodriguez_2020,Ferraro_2020,Ferraro_2021}. These differences lead for example to the electronic Hong-Ou-Mandel effect, where electrons arriving simultaneously on each side of a quantum point contact anti-bunch due to destructive two-particle interference and leave via different output arms (in contrast to the bunching of photons due to constructive interference)~\cite{Bocquillon_2013,Freulon_2015,Marguerite_2016,Glattli_2016,Glattli_2016b,Marguerite_2016b}. Experiments have also demonstrated tomography of single electrons~\cite{Grenier_2011,Jullien_2014,Fletcher_2019,Bisognin_2019}, measured the time of flight~\cite{Kataoka_2013,Waldie_2015,Kataoka_2016,Roussely_2018} and counting statistics~\cite{Fricke_2013,Ubbelohde_2014} of injected particles, and have emitted particles well above the Fermi level~\cite{Johnson_2018,Rodriguez_2019}.

All of these experiments have stimulated a wide range of theoretical activities. Several works have explored the possibilities of generating entanglement using dynamic single-electron sources~\cite{Samuelsson_2004,Blatter_2005,Beenakker_2005,Samuelsson_2005,Hofer_2013,Dasenbrook_2015,Dasenbrook_2015b,Dasenbrook_2016,Hofer_2017,Samuelsson_2020}. Heat transport and fluctuations of dynamic single-electron emitters have also been considered~\cite{Moskalets_2009,Battista_2013,Moskalets_2016,Samuelsson_2017,Samuelsson_2019,Ferraro_2019} as well as the distribution of waiting times between emitted particles~\cite{Dasenbrook_2014,Dasenbrook_2016b,Mi_2018,Burset_2019,Brange_2021}. In addition, methods for performing signal processing of quantum electric currents have been developed~\cite{Roussel_2017,Roussel_2021}, and combinations of voltage pulses and superconductors have been discussed~\cite{Ferraro_2019b,Averin_2020}. The combined effects of several single-particle sources as well as multi-particle emitters have also been investigated~\cite{Splettstoesser_2008,Splettstoesser_2009,Moskalets_2011,Juergens_2011,Sim_2016,Misiorny_2018,Ferraro_2018b,Moskalets_2020}. At the heart of electron quantum optics lies the Landauer-Büttiker formalism, which describes the transmission and reflection of incoming particles on a mesoscopic conductor in terms of scattering matrices~\cite{Buttiker_1992}. Originally, it was formulated for static setups, however, it can be extended to dynamically driven systems using Floquet scattering theory, which accounts for the exchange of energy quanta between particles and an external classical driving field \cite{Moskalets_2002,Moskalets_2008,Moskalets_2013,Moskalets_2014,Moskalets_2015b,Moskalets_2017,Moskalets_2017b,Moskalets_book}.

In this article, we investigate theoretically the interference of multi-particle excitations emitted into an electronic Mach-Zehnder interferometer building on earlier works on either static voltages~\cite{Chung_2005}, periodically modulated interferometers \cite{Chung_2007}, or dynamic charge emitters \cite{Hofer_2014}. We focus in particular on the injection of clean multi-particle pulses into the interferometer, and we show how the visibility measured in the outputs can be related to the excess correlation function of the incoming pulse, which can be further decomposed into elementary contributions from the individual charges making up the pulse. Based on this understanding, we can interpret an observed Fraunhofer-like diffraction pattern as arising due to the interference of the excess correlation functions of various elementary single-electron components of the multi-particle pulse. These findings may be observed in future experiments on electron quantum optics with dynamic charge emitters and electronic Mach-Zehnder interferometers.

The rest of the paper is organized as follows: In Sec.~\ref{sec:theory}, we review the theoretical description of quantum transport in dynamically driven mesoscopic conductors based on the Floquet scattering formalism, and we introduce the notion of  correlation functions in electron quantum optics. In Sec.~\ref{sec:MZI}, we describe the electronic Mach-Zehnder interferometer and show how the current in the outputs can be divided into a classical contribution and an interference term, which vanishes as the temperature is increased. We also show how the visibility of the interferometer can be expressed in terms of the excess correlation function of the injected charge pulses. Section~\ref{sec:visi} is devoted to our results and analysis of the visibility of the interferometer, which we interpret in terms of  interferences between the different single-particle components of the multi-charge pulse, giving rise to a Fraunhofer-like diffraction pattern. Finally, in Sec.~\ref{sec:conc}, we describe the main conclusions of our work.
 
\section{Theoretical Background} 
\label{sec:theory}

\subsection{Floquet scattering theory}

Mesoscopic conductors driven by a periodic voltage can be described using Floquet scattering theory~\cite{Moskalets_book}. We thus consider systems governed by a time-periodic Hamiltonian $H(x, t) = H(x, t + \mathcal{T})$ with period $\mathcal{T}$ and frequency $\Omega = 2\pi/\mathcal{T}$. According to the Floquet theorem, a complete set of solutions to the time-dependent Schr{\"o}dinger equation can be written as \cite{Moskalets_book}
\begin{equation}
	\Psi_\ell(x, t) = \e^{-\frac{iE_\ell t}{\hbar}}\phi_\ell(x, t), \quad
	\phi_\ell(x, t) = \phi_\ell(x, t + \mathcal{T}).
\end{equation}
We can expand the periodic function $\phi_\ell(x, t)$ in a Fourier series, such that the wave functions become
\begin{equation}
	\Psi_\ell(x, t) = \e^{-\frac{iE_\ell t}{\hbar}}\sum_{n=-\infty}^{\infty}\e^{-in\Omega t}\phi_{\ell,n}(x).
	\label{eq:floquet_wave_function}
\end{equation}
The wave functions are invariant with respect to the shift $E_\ell \rightarrow E_\ell + m\hbar\Omega$, with $m\in\mathbb{Z}$. Therefore, the energy $E_\ell$ is a quasi-energy (Floquet energy), which is defined only up to integer multiples of $\hbar\Omega$. 

\begin{figure}
	\centering
	\includegraphics[width=0.9\textwidth]{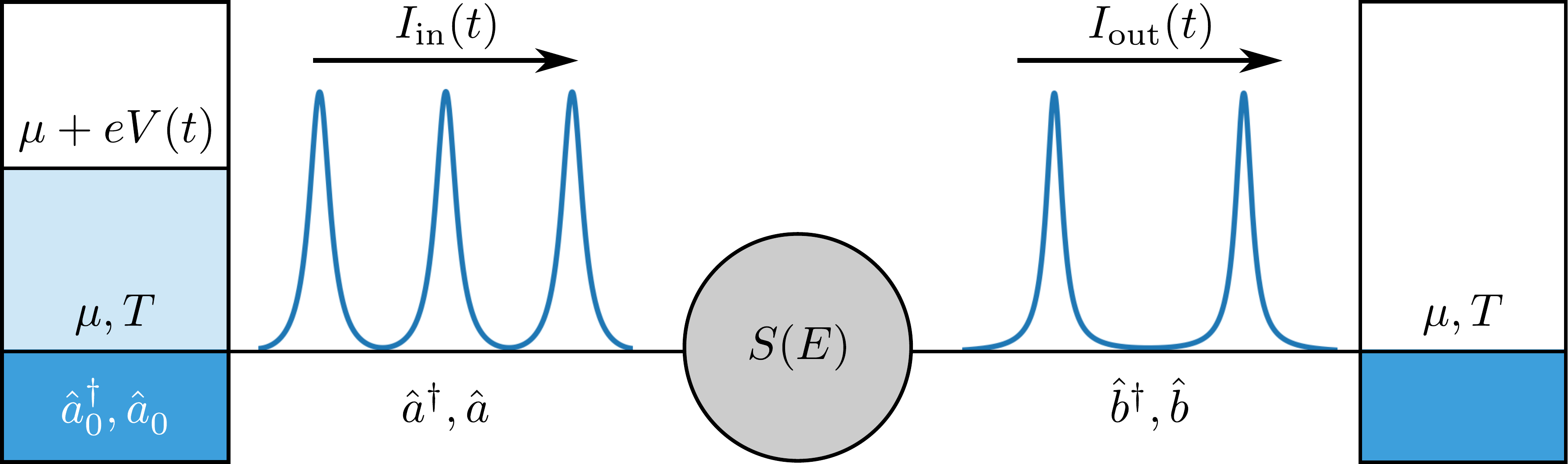}
	\caption{Illustration of a periodically-driven mesoscopic scatterer. Periodic voltage pulses, $V(t)$, are applied to the left electrode, so that electrons are emitted towards the central scatterer, described by the scattering matrix $S(E)$. There, they are either reflected back into the source electrode or transmitted to the right drain electrode, where the electric current is measured. The fermionic operators $\hat{a}^\dagger_0$ and $\hat{a}_0$ describe electrons in equilibrium at temperature $T$ and chemical potential $\mu$, before the application of the voltage pulses. The operators $\hat{a}^\dagger$ and $\hat{a}$ describe electrons after the application of the voltage pulses, while the operators $\hat{b}^\dagger$ and $\hat{b}$ describe electrons that have been transmitted through the scatterer. The different operators are related via the Floquet scattering matrix according to Eqs.~(\ref{eq:floquet_operators_incident}) and (\ref{eq:scattered_operators}). The electric current before and after the scatterer are denoted as $I_{\text{in}}(t)$ and $I_{\text{out}}(t)$, respectively.
	}
	\label{fig:driven_system_schematic}	
\end{figure}

In the following, we consider a mesoscopic scatterer that is connected by ballistic leads to two electronic reservoirs, and a periodic voltage $V(t) = V(t + \mathcal{T})$ is applied to one of them, see Fig.~\ref{fig:driven_system_schematic}. Electrons are then emitted from the driven reservoir towards the scatterer from where they are either reflected back into the source electrode, or they are transmitted through the scatterer and eventually reach the drain electrode. An applied voltage can be treated as a spatially uniform potential~\cite{Moskalets_book}. The wave function of  electrons leaving the reservoir can then be derived from the Schrödinger equation, 
\begin{equation}
	\Psi(x, t)=\e^{-\frac{iEt}{\hbar}}\psi_0(x)J(t),
	\label{eq:wave_function_driven_start} 
\end{equation}
where $\psi_0(x)$ is the solution without the voltage pulses, and $J(t)$ is a time-dependent phase factor,
\begin{equation} 
	J(t) =  \e^ {-ie\int\limits_{-\infty}^{t}V(t')dt'/\hbar}.
	\label{eq:j_t}
\end{equation}
Owing to the periodicity of the time-dependent voltage, we can expand $J(t)$ in a Fourier series as
\begin{equation}
	J(t) = \sum_{n = -\infty}^{\infty}J_n\e^{-in\Omega t}, 
	\label{eq:j_fourier}
\end{equation}
with the Fourier coefficients given as
\begin{equation}
	J_n = \int\limits_{0}^{\mathcal{T}}\frac{dt}{\mathcal{T}}\e^{in\Omega t}J(t).
	\label{eq:j_fourier-coef}
\end{equation}
As a result, the wave function of electrons leaving the driven contact becomes
\begin{equation}
	\Psi(x, t)=\e^{-\frac{iEt}{\hbar}}\sum_{n = -\infty}^{\infty}\e^{-in\Omega t}J_n\psi_0(x),
	\label{eq:wave_function_driven_final}
\end{equation}
which is obviously a Floquet wave function as in Eq.~(\ref{eq:floquet_wave_function}). Now, based on Eq.~(\ref{eq:wave_function_driven_final}), the possible energies of an electron after being excited by the periodic voltage are $E_n = E + n\hbar\Omega$, where $n$ is a positive or negative integer and $E$ is the energy of the electron before interacting with the voltage pulses.

In what follows, we consider quasi-one-dimensional transport as in the chiral edge states of a quantum Hall system. We assume that the propagating electrons have energies close to the Fermi energy~$\mu$, and approximate their dispersion relation as linear, $E - \mu\simeq\hbar v_F(k - k_F)$, where $k_F$ is the Fermi wavevector. Thus, the electrons are assumed to propagate at the Fermi velocity $v_F$. 
Using second quantization, the operators $\hat{a}^\dagger_{0,j}(E)$ and $\hat{a}^{}_{0,j}(E)$ describe electrons with energy $E$ in the reservoir labeled~$j$ connected to the incoming lead, before being excited by the voltage pulses. These electrons are in thermal equilibrium, allowing us to write their quantum-statistical average as 
\begin{equation}
	\langle \hat{a}^\dagger_{0,i}(E)\hat{a}_{0,j}(E') \rangle \equiv \langle 0| \hat{a}^\dagger_{0,i}(E)\hat{a}_{0,j}(E') |0 \rangle = \delta_{ij}\delta(E - E')f_j(E),
	\label{eq:equilibrium_average}
\end{equation}
where $|0\rangle$ is the state of the undisturbed Fermi sea, and $f_j(E)=1/(\e^{\beta(E-\mu)}+1)$ is the Fermi-Dirac distribution with $\beta = 1/(k_BT)$ being the inverse electronic temperature. 
We also define the creation and annihilation operators $\hat{a}_j^\dagger$ and $\hat{a}_j$ for electrons incident on the scatterer from lead $j$, after being excited by the voltage. They are related to the equilibrium ones as
\begin{equation}
	\hat{a}_j(E) = \sum_{n = -\infty}^{\infty} J_n \hat{a}_{0,j}(E_{-n}), \quad
	\hat{a}_j^\dagger(E) = \sum_{n = -\infty}^{\infty} J_n^* \hat{a}^\dagger_{0,j}(E_{-n}). 
	\label{eq:floquet_operators_incident}
\end{equation}
Next, we describe the transmission and reflection of charge pulses on the central scatterer. The operators $b_i^\dagger(E)$ and $b_i(E)$ for scattered electrons, with $i$ labeling the reservoirs, are related to the incident ones as 
\begin{equation}
	\hat{b}_i(E) = \sum_{j}S_{ij}(E)\hat{a}_j(E), \quad
	\hat{b}_i^\dagger(E) = \sum_{j}S_{ij}^*(E)\hat{a}_j^\dagger(E),
	\label{eq:scattered_operators}
\end{equation}
where $S(E)$ is the scattering matrix. 
Then, by combining Eqs.~(\ref{eq:floquet_operators_incident}) and (\ref{eq:scattered_operators}), we can express the operators for the scattered electrons as
\begin{equation}
	\hat{b}_i(E) = \sum_{j}\sum_{n = -\infty}^{\infty} S_{ij}(E)J_n \hat{a}_{0,j}(E_{-n}), \quad
	\hat{b}_i^\dagger(E) = \sum_{j}\sum_{n = -\infty}^{\infty} S_{ij}^*(E)J_n^* \hat{a}^\dagger_{0,j}(E_{-n}),
	\label{eq:floquet_operators_scattered}
\end{equation}
whereby we can identify the product $S(E_n)J_n$ as the Floquet scattering matrix, 
\begin{equation}
	S_F(E_n, E) = S(E_n)J_n,
	\label{eq:floquet_scattering_matrix}
\end{equation}
for a static conductor driven by periodic voltage pulses. 
The Floquet scattering matrix contains the probability amplitudes for an electron with energy $E$ to change its energy to $E_n=E+n\hbar \Omega$ by exchanging $n$ modulation quanta of size $\hbar \Omega$ with the voltage and to be transmitted through the scatterer. For our purposes, it is useful that the effects of the voltage drive and the scatterer can be factorized as in Eq.~(\ref{eq:floquet_scattering_matrix}), which would not be the case if the scatterer itself was modulated in time.

Finally, the electric current generated by a time-dependent voltage reads~\cite{Moskalets_book,Buttiker_1992,Moskalets_2007}
\begin{equation}
	I_i(t) = \frac{e}{h}\int\limits_{0}^{\infty}\int\limits_{0}^{\infty}dEdE'\e^{\frac{i}{\hbar}(E-E')t} \langle \hat{b}_i^\dagger(E)\hat{b}_i(E') - \hat{a}_i^\dagger(E)\hat{a}_i(E')\rangle .
	\label{eq:current_general}	
\end{equation}
Using Eqs.~(\ref{eq:equilibrium_average}) and (\ref{eq:floquet_operators_scattered}), the time-dependent current can be recast as 
\begin{equation}
	I_i(t) = \frac{e}{h}\int\limits_{0}^{\infty}dE \sum_j \sum_{n = -\infty}^{\infty}\sum_{m = -\infty}^{\infty}[f_j(E)-f_i(E_n)]\e^{-i\Omega t(m-n)}J_n^*J_{m}S_{ij}^*(E_n)S_{ij}(E_{m}),
	\label{eq:current_floquet_js}
\end{equation}
where the Fourier components of the voltage pulses are defined in Eq.~(\ref{eq:j_fourier-coef}). 

\subsection{Voltage pulses}

We now specify the types of voltage pulses that we will consider. A current that carries the charge of one electron per period can be created by applying a periodic voltage $V(t)$, which satisfies the condition $\frac{e^2}{h}\int_t^{t+\mathcal{T}} V(t')dt' = e$, where $\mathcal{T}$ is the period of the drive~\cite{Keeling_2006}. For almost any voltage drive, such a current is accompanied by neutral electron-hole pairs, and the resulting charge pulses consist of more than just a single electron. However, by applying Lorentzian-shaped pulses, clean single-electron excitations can be created \cite{Levitov_1996,Levitov_1997, Keeling_2006}. Experimentally, such single-electron pulses were first realized by Dubois and coworkers, who named them levitons~\cite{Dubois_2013,Dubois_2013b}.
A Lorentzian voltage pulse has the form
\begin{equation}
	V(t) = \bar{n}\frac{\hbar}{e}\frac{2\Gamma}{t^2+\Gamma^2},
	\label{eq:lorentzian_voltage_pulse}
\end{equation}
where $\Gamma$ is the half-width of the pulse,  and $\bar{n}$ controls the average charge per voltage pulse, which can be either integer or non-integer. Technically, we treat single voltage pulses, or a train of a finite number of pulses, within the Floquet scattering formalism by considering a period that is much longer than all other time scales in the problem.  We will also consider single Gaussian voltage pulses of the form
\begin{equation}
	V(t) = \bar{n}\frac{\hbar}{e}\frac{\sqrt{4\pi\log(2)}}{\Gamma}\exp\left(-\frac{\log(2)t^2}{\Gamma^2}\right),
	\label{eq:gaussian_voltage_pulse}
\end{equation}
as well as sinusoidal voltage pulses reading
\begin{equation}
	V(t) = 
	\begin{cases}
		\bar{n}\frac{\hbar}{e}\frac{\pi}{2\Gamma}(\sin(\frac{\pi t}{2\Gamma} + \frac{\pi}{2}) + 1), \quad \text{if} \quad |t| \leq \Gamma \\
		0, \quad \text{otherwise}
	\end{cases}
	\label{eq:sinusoidal_voltage_pulse}    
\end{equation}
Finally, a sequence of $m$ pulses is given as
\begin{equation}\label{eq:finite-train}
	V(t) = \sum_{j=0}^{m-1} V_1(t-jw),
\end{equation}
where $V_1$ is the voltage for a single pulse, and $w$ is the separation between the pulses. 

\subsection{Excess correlation function} 

Electron quantum optics experiments aim to realize ideas from quantum optics, however, with electrons in solid-state architectures instead of photons in vacuum or in optical fibres. Therefore, it is relevant to adapt the theoretical toolbox from quantum optics to condensed matter systems. Among the useful tools in quantum optics are the correlation functions introduced by Glauber~\cite{Glauber1963photon,Glauber1963quantum,Glauber1963coherence}. Correlation functions, also known as Green's functions, involve quantum statistical averages of products of field operators. To facilitate the analysis of electron quantum optics experiments, the ideas of Glauber have recently been extended to electronic systems and mesoscopic conductors~\cite{Grenier_2011,Grenier_2011b,Haack_2013}. 
An important difference in the electronic case compared to quantum optics is that, unlike photons which often propagate in vacuum, electrons in mesoscopic conductors propagate on top of the Fermi sea. For this reason, we need to use excess correlation functions, which subtract the effects of the underlying Fermi sea, leaving only correlations between the propagating electrons close to the Fermi level.  

The first-order excess correlation function $\tilde{G}^{(1)}$ contains all information about the single-particle states of the emitted electrons~\cite{Haack_2013}. It also provides some information about multi-particle states, although their full characterization requires higher-order correlation functions~\cite{Moskalets_2018}.  For electrons injected by a voltage drive, the first-order excess correlation function is defined as~\cite{Grenier_2011}
\begin{equation}
	\tilde{G}_{ij}^{(1)}(t_1, t_2) = \langle \hat{\Psi}_{i}^\dagger(t_1)\hat{\Psi}_{j}(t_2)\rangle_{\text{on}} - \langle \hat{\Psi}_{i}^\dagger(t_1)\hat{\Psi}_{j}(t_2) \rangle_{\text{off}},
	\label{eq:G1_def}
\end{equation}
where the term with the subscript "on" contains field operators for the case with a time-dependent voltage $V(t)$, and the term with the subscript "off" corresponds to the case where there is no voltage drive. Due to the linear dispersion relation, we can evaluate the correlation function anywhere along a lead, and we generally leave out the spatial dependence in these definitions as well as in wave functions and other related quantities. If the injected electrons are transmitted through a scatterer, we must separately define the correlation functions before and after the scatterer. However, we will focus on the correlation function before the scatterer,  which can be related to measurable quantities such as  the interference current and the visibility in a Mach-Zehnder interferometer as we will see. 

To evaluate the correlation function, we first define the electronic field operators that create an electron at position $x$ in  lead $j$ as
\begin{equation}
	\hat{\Psi}_j^\dagger(x, t) = \frac{1}{\sqrt{hv_F}}\int_{0}^{\infty}\left[ \hat{a}_j^\dagger(E)\psi_{E,j}^{(\text{in})*}(x, t) + \hat{b}_j^\dagger(E)\psi_{E,j}^{(\text{out})*}(x, t) \right] dE,
	\label{eq:field_operator_definition}
\end{equation}
where $\psi_{E,j}^{(\text{in}/\text{out})}(x, t)/\sqrt{hv_F}$ are the incident and scattered wave functions for electrons in lead $j$ with energy $E$. The prefactor $1/\sqrt{hv_F}$ ensures that the wave functions carry a unit flux of particles. 

Without a voltage applied to the reservoir, the creation and annihilation operators in the lead are simply the same as the reservoir operators, $\hat{a}^\dagger_{0,j}$, $\hat{a}^{}_{0,j}$. However, when a time-dependent voltage is applied to the reservoir, the operators in the lead are related to the reservoir operators according to Eq.~(\ref{eq:floquet_operators_incident}). 
To keep the notation simple, we only consider the correlation function for operators in the same lead and henceforth omit the indices of the leads. 
The correlation function in Eq.~(\ref{eq:G1_def}) then reads
\begin{equation}
	\begin{aligned}
		\tilde{G}^{(1)}(t_1, t_2) =& \left\langle \frac{1}{hv_F} \int_{0}^{\infty} \hat{a}^\dagger(E)\psi_{E}^{(in)*}(t_1)dE \int_{0}^{\infty} \hat{a}(E')\psi_{E'}^{(in)}(t_2)dE' \right\rangle
		\\
		-& \left\langle \frac{1}{hv_F} \int_{0}^{\infty} \hat{a}^\dagger_0(E)\psi_{E}^{(in)*}(t_1)dE \int_{0}^{\infty} \hat{a}^{}_0(E')\psi_{E'}^{(in)}(t_2)dE' \right\rangle.
		\label{eq:G1_1}
	\end{aligned}
\end{equation}
Since the wave functions in the lead are plane waves, $\psi_E^{(in)}(x, t) = \e^{ikx} \e^{- iEt/\hbar}$, we moreover find
\begin{equation}
		\tilde{G}^{(1)}(t_1, t_2) =  \frac{1}{hv_F} \int_{0}^{\infty}\int_{0}^{\infty}dE'dE \e^{i(Et_1 - E't_2)/\hbar}  \left[ \left\langle \hat{a}^\dagger(E) \hat{a}(E') \right\rangle - \left\langle \hat{a}^\dagger_0(E) \hat{a}^{}_0(E') \right\rangle \right].
		\label{eq:G1_2}
\end{equation}  
Using Eq.~(\ref{eq:floquet_operators_incident}), we can insert equilibrium operators in this expression, which allows us to evaluate the quantum-statistical averages based on Eq.~(\ref{eq:equilibrium_average}). We can perform one of the integrations, which yields
\begin{equation}
	\tilde{G}^{(1)}(t_1, t_2) = \frac{1}{hv_F} \int_{0}^{\infty}dE \e^{iE(t_1-t_2)/\hbar} \left[ \sum_{n, m = -\infty}^{\infty}J_n^*J_m \e^{i(n-m)\Omega t_1}f(E_{-m}) - f(E) \right].
	\label{eq:G1_3}
\end{equation}  
This expression holds for any temperature and driving frequency. However, we still need to evaluate the last integral. To this end, we employ an approximation which is valid at low temperatures and driving frequencies compared to the Fermi energy, which is reasonable for mesoscopic conductors.

Generally, we need to evaluate energy integrals of the form
\begin{equation}
	\int_{0}^{\infty}f(E_n)\e^{ia(E-\mu)}dE = \int_{0}^{\infty}\frac{\e^{ia(E-\mu)}}{1 + \e^{\beta(E+n\hbar\Omega-\mu)}}dE, \quad a\in\mathbb{R}.
	\label{eq:energy_integral_1}
\end{equation}
Changing the integration variable to $\varepsilon = \beta(E+n\hbar\Omega - \mu)$ allows us to rewrite the integral as
\begin{equation}
\int_{\beta(n\hbar\Omega-\mu)}^{\infty}\frac{\e^{ia(\epsilon/\beta-n\hbar\Omega)}}{1 + \e^{\epsilon}} \frac{d\epsilon}{\beta} 
	\simeq \frac{\e^{-ian\hbar\Omega}}{\beta}\int_{-\infty}^{\infty}\frac{\e^{ia\varepsilon/\beta}}{1 + \e^{\varepsilon}}d\varepsilon,
	\label{eq:energy_integral_2}
\end{equation}
where we have assumed that both the temperature and the driving frequency are small compared to the Fermi energy, namely, $\mu\beta\gg1$ and $\mu\gg n\hbar\Omega$, where $n$ is among the largest numbers of excitation quanta that an electron can emit or absorb because of the time-dependent voltage. As a result, we can extend the lower limit of the integral to minus infinity and evaluate the resulting integral using contour integration in the complex plane, leading to the simple expression 
\begin{equation}
	\int_{0}^{\infty}f(E_n)\e^{ia(E-\mu)}dE \simeq \frac{\e^{-ian\hbar\Omega}}{ia}\chi\left(\frac{\pi a}{\beta}\right), \quad \chi(x) \equiv \frac{x}{\sinh(x)}.
	\label{eq:energy_integral_3}
\end{equation}
Finally, by combining Eq.~(\ref{eq:G1_3}) and Eq.~(\ref{eq:energy_integral_3}), we arrive at the useful approximation
\begin{equation}
	\tilde{G}^{(1)}(t_1, t_2) \simeq \frac{\e^{i\mu(t_1-t_2)/\hbar}}{2\pi i(t_1-t_2)v_F}\chi\left(\frac{\pi(t_1-t_2)}{\hbar\beta}\right) \left[ \sum_{n,m=-\infty}^{\infty}J_n^*J_m\e^{i\Omega(nt_1-mt_2)} - 1 \right].
	\label{eq:G1_4}
\end{equation}

To proceed, we note that the sums in Eq.~(\ref{eq:G1_4}) exactly correspond to  the Fourier expansions of $J^*(t_1)$ and $J(t_2)$ in Eq.~(\ref{eq:j_fourier}). Moreover, by using Eq.~(\ref{eq:j_t}), we find    
\begin{equation}
	J^*(t_1)J(t_2) = \e^{-i\Phi(t_1, t_2)},\,\,\,\,\,
	\Phi(t_1, t_2) = \frac{e}{\hbar}\int_{t_1}^{t_2}V(t)dt,
\label{eq:j_j_star}
\end{equation} 
such that the correlation function can be written as
\begin{equation}
	\tilde{G}^{(1)}(t_1, t_2) \simeq \frac{\e^{i\mu(t_1 - t_2)/\hbar}}{v_F} \chi\left( \frac{\pi(t_1-t_2)}{\hbar\beta} \right) G^{(1)}(t_1, t_2),
	\label{eq:G1_5}
\end{equation}
where we have introduced the envelope correlation function~\cite{Moskalets_2015}
\begin{equation}
	G^{(1)}(t_1, t_2) = \frac{\e^{-i\Phi(t_1, t_2)} - 1}{2\pi i(t_1 - t_2)}.
	\label{eq:G1_envelope}
\end{equation}
The form of the correlation function in Eq.~(\ref{eq:G1_5})  is particularly useful, as it separates the effects of the Fermi energy, the temperature, and the voltage~\cite{Grenier_2013,Glattli_2016b}. The effects of the Fermi energy enter as the phase factor $\e^{i\mu(t_1 - t_2)/\hbar}$, while the temperature enters via the function $\chi$, which decreases from $\chi = 1$ at zero temperature to $\chi\simeq0$ at large temperatures, where quantum coherence is lost. Finally, the effects of the time-dependent voltage are contained in $G^{(1)}$ through the phase factor $\Phi(t_1, t_2)$. It is easy to add the other factors, if needed, and we therefore focus on the envelope correlation function.

\section{Mach-Zehnder interferometer}
\label{sec:MZI}

\begin{figure}
    \centering
	\includegraphics[width=0.8\textwidth]{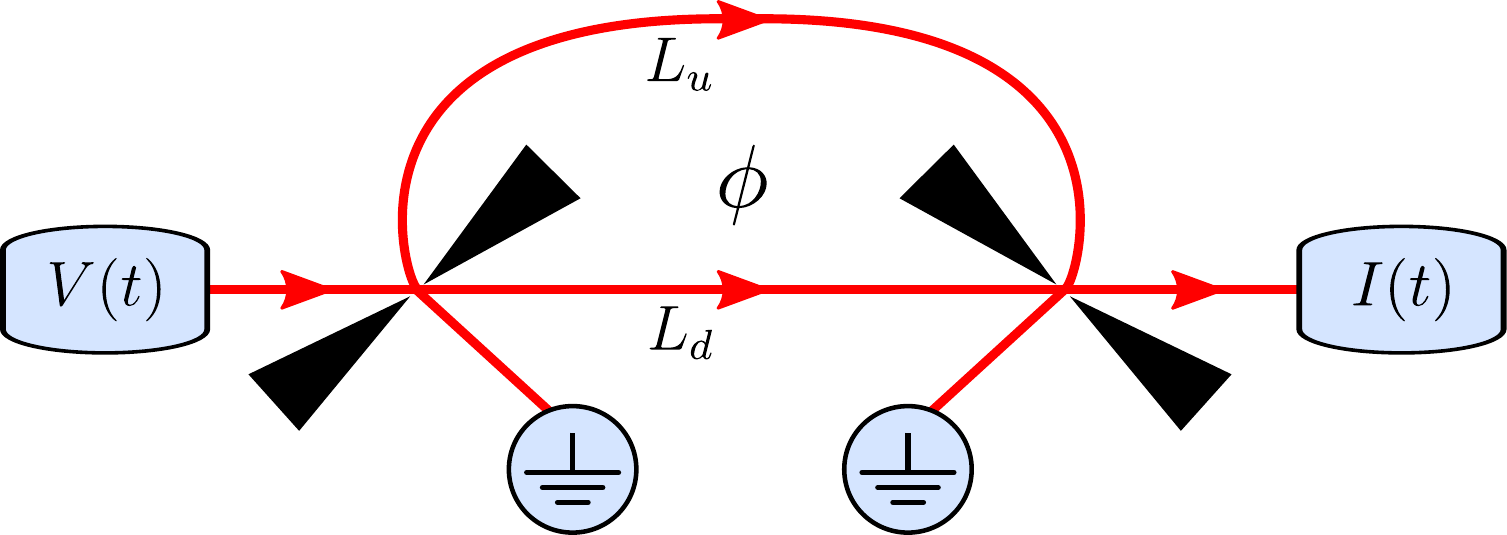}
	\caption{
		Schematic of an electronic Mach-Zehnder interferometer implemented with chiral edge states. Charges are emitted in the upper input by the application of a time-dependent voltage $V(t)$. Incoming particles are coherently split at the first quantum point contact, which acts as a beam splitter, and they then propagate along the upper or lower arms of the interferometer of lengths $L_u$ and $L_d$, respectively. The particles recombine at the second quantum point contact and then leave via the upper or lower output of the interferometer. The injected particles propagate at the Fermi velocity, $v_F$, such that it takes the time $\tau_{u,d}=L_{u,d}/v_F$ to travel along the upper or lower arm. In addition, the arms enclose a magnetic flux, which causes a phase shift $\phi=\phi_u-\phi_d$ between electrons traveling along each of the arms. We consider the time-dependent current in the upper output of the interferometer, $I(t)$.
	}
	\label{fig:mzi_schematic}
\end{figure}

An electronic Mach-Zehnder interferometer can be used to investigate interference effects of electrons, similarly to an optical Mach-Zehnder interferometer for photons. The setup can be realized using chiral edge states in a Corbino disk geometry \cite{Ji_2003}. The interferometer is made up of two quantum point contacts that act as beam splitters with energy independent reflection and transmission amplitudes that we denote by $r_1, r_2$ and $t_1, t_2$. There  are two paths for electrons to travel from the first beam splitter to the next one, as illustrated in Fig.~\ref{fig:mzi_schematic}, and we denote the length of the upper and lower paths as $L_u$ and $L_d$, respectively. In addition, a magnetic flux threads the area enclosed by the paths, causing electrons to acquire the additional phases $\phi_u$ and $\phi_d$ depending on the path. In the following, we analyze the current injected in the upper input and the current that reaches the upper output.

\subsection{Injected current}
We start by evaluating the time-dependent current injected in the input. In this case, the scattering matrix in Eq.~(\ref{eq:current_floquet_js}) equals unity, and the integral over energy becomes $n\hbar\Omega$. We then find
\begin{equation}
I_{\text{in}}(t) = \frac{e}{2\pi}\sum_{n=-\infty}^{\infty}n\Omega J_n^*\e^{i\Omega tn}\sum_{m=-\infty}^{\infty}J_m\e^{-i\Omega tm}.
\end{equation}
From Eq.(\ref{eq:j_fourier}), we recognize that this expression can be rewritten as
\begin{equation}
I_{\text{in}}(t) = \frac{e}{2\pi}\left(-i\frac{dJ^*(t)}{dt}\right)J(t).
\end{equation}
Moreover, using that 
\begin{equation}
	-i\frac{dJ^*(t)}{dt}J(t) = \frac{e}{\hbar}V(t),
\end{equation}
we arrive at the simple expression for the injected current 
\begin{equation}\label{eq:inj-curr}
I_{\text{in}}(t) = \frac{e^2}{h}V(t),
\end{equation}
showing that it is given simply by the conductance quantum, $G_0=e^2/h$, times the the applied voltage. 

\subsection{Output current}

Next, we turn to the current in the upper output of the interferometer. The scattering amplitude is given by \cite{Ji_2003,Haack_2014}
\begin{equation}
S(E) = r_1r_2\e^{i(L_uk_F+\tau_u(E-\mu)/\hbar+\phi_u)} + t_1t_2\e^{i(L_dk_F+\tau_d(E-\mu)/\hbar+\phi_d)},
\label{eq:mzi_scattering_amplitude}
\end{equation}
where $\tau_{u,d} = L_{u,d}/v_F$ are the times it takes for electrons to travel along the upper or lower arm. The two contributions above describe the propagation of electrons in each of the two arms. We now see that the product of scattering matrices can be written as a sum of "classical" and "interference" contributions,
\begin{equation}
S^*(E_n)S(E_m) = T_{\text{cl}} + T_{\text{int}}, 
\label{eq:transcont}
\end{equation}
with
\begin{equation}
T_{\text{cl}} = R_1R_2\e^{i(m-n)\Omega\tau_u} + T_1T_2\e^{i(m-n)\Omega\tau_d} 
\end{equation}
and
\begin{equation}
T_{\text{int}} = -\sqrt{R_1R_2T_1T_2}\left( \e^{i[ (k_F v_F + (E-\mu)/\hbar) \Delta\tau  + (m\tau_u-n\tau_d)\Omega-\phi]} + \e^{-i[ (k_F v_F + (E-\mu) /\hbar) \Delta\tau + (n\tau_u-m\tau_d)\Omega-\phi]} \right),
\label{eq:mzi_scattering_amplitude_parts}
\end{equation}
where $\Delta\tau =\tau_u-\tau_d$ is the time difference between traveling along each arm, and $\phi=\phi_u-\phi_d$ is the phase difference controlled by the magnetic flux. We have also used the unitarity of the scattering matrices of the two quantum point contacts to deduce that $r_jt_j^*=-r_j^*t_j$, which leads to $r_1r_2t_1^*t_2^*=r_1^*r_2^*t_1t_2=-\sqrt{R_1R_2T_1T_2}$, where $T_j=|t_j|^2$ and $R_j=1-T_j$, $j=1,2$. 

Based on Eq.~(\ref{eq:current_floquet_js}), we see that the current in the output can also be divided into a classical part and an interference contribution. Specifically, we find
\begin{equation}
I_{\text{out}}(t) = I_{\text{cl}}(t) + I_{\text{int}}(t),  
\end{equation}
with the two contributions given by the corresponding terms of the transmission in Eq.~(\ref{eq:transcont}), 
\begin{equation}
I_{\text{cl}(\text{int})}(t) = \frac{e}{h}\sum_{n = -\infty}^{\infty}\sum_{m = -\infty}^{\infty}\e^{-i\Omega t(m-n)}J_n^*J_{m} \int\limits_{0}^{\infty} [f(E)-f(E_n)]T_{\text{cl}(\text{int})}dE.
\end{equation}
For the classical part $I_\text{cl}$, there is no energy dependence in the scattering amplitudes, so $I_\text{cl}$ can be derived in essentially the same way as the injected current in Eq.~(\ref{eq:inj-curr}) with the only differences being the prefactors $R_1R_2$ and $T_1T_2$ and the exponential factors containing $\tau_u$ and $\tau_d$. 
Hence, we find
\begin{equation}
	I_{\text{cl}}(t) = \frac{e^2}{h}\left[R_1R_2V(t - \tau_u) + T_1T_2V(t - \tau_d) \right],	
	\label{eq:mzi_current_d}
\end{equation}
which is simply a sum of shifted and scaled injected currents, and this result can be understood without resorting to quantum mechanics. By contrast, quantum effects are important for the interference current. For $\Delta\tau=0$, the scattering amplitudes are energy independent, and we find  the interference current just as we found the classical one. On the other hand, for $\Delta\tau\neq0$, the scattering amplitudes are energy dependent, and we find the interference curret using Eq.~(\ref{eq:energy_integral_3}). The interference current then becomes
\begin{equation}\label{eq:mzi_current_nd}
	I_{\text{int}}(t) = -2e\sqrt{R_1R_2T_1T_2}
	\begin{cases}
			\frac{e}{h} V(t-\tau_{u/d})\cos\phi, & \Delta\tau = 0\\
		 \chi\left(\frac{\pi\Delta\tau}{\hbar\beta}\right) \mathrm{Re} \left\{ \e^{-i(k_Fv_F\Delta\tau-\phi)} G^{(1)}(t-\tau_u,t-\tau_d) \right\}, & \Delta\tau \neq 0
	\end{cases} 
\end{equation}
with the envelope correlation function $G^{(1)}$ defined in Eq.~(\ref{eq:G1_envelope}), and the function $\chi$ given in Eq.~(\ref{eq:energy_integral_3}). 

\subsection{Transferred charge}

In addition to the output current, we can consider the transferred charge given by the integral
\begin{equation}
	Q = \int I_{\text{out}}(t) dt,
	\label{eq:charge}
\end{equation}
which is taken over a period, which for a finite number of pulses is assumed to be much longer than any other time scale in the problem. The transferred charge can again be separated into classical and interference contributions as
\begin{equation}
Q = Q_\text{cl} + Q_\text{int},
\label{eq:transmitted_charge}
\end{equation}
where $Q_\text{cl}$ and $Q_\text{int}$ are given by Eq.~(\ref{eq:charge}) by inserting the corresponding currents.

From Eq.~(\ref{eq:mzi_current_d}), we see that the classical contribution reads 
\begin{equation}
	Q_\text{cl} = \int I_\text{cl}(t)dt = \frac{e^2}{h}\left(R_1R_2 + T_1T_2 \right)\int V(t)dt
	\label{eq:Q_cl}
\end{equation}
which is simply the charge of the injected current multiplied by $R_1R_2 + T_1T_2$. The charge transferred due to the interference current follows from Eq.~(\ref{eq:mzi_current_nd}) and can be expressed as
\begin{equation}
	Q_\text{int} = \int I_\text{int}(t)dt = -2e\sqrt{R_1R_2T_1T_2} \chi\left(\frac{\pi\Delta\tau}{\hbar\beta}\right) \text{Re}\left\{\e^{-i(k_Fv_F\Delta\tau-\phi)}\int G^{(1)}(t-\tau_u, t-\tau_d) dt \right\}
	\label{eq:Q_int}
\end{equation}
in terms of the excess correlation function, where we have assumed that $\Delta\tau\neq0$.

\subsection{Visibility}
A common way to characterize interference is to measure the  visibility of a physical observable. This technique can be applied to various situations involving wave interference, from classical electromagnetic waves to quantum mechanical particles. Typically, there is a physical observable which oscillates as a function of a control parameter. The usual definition of visibility in this context is the ratio of the amplitude of the oscillations over the average value of the oscillations. The charge visibility has so far been the main method for characterizing interference effects in Mach-Zehnder interferometers~\cite{Ji_2003}.

The charge transferred through a Mach-Zehnder interferometer oscillates as a function of the magnetic flux according to Eqs.~(\ref{eq:charge}, \ref{eq:Q_cl}, \ref{eq:Q_int}). The standard definition of the charge visibility is then 
\begin{equation}
	\nu = \frac{\max[Q(\phi)] - \min[Q(\phi)]}{\max[Q(\phi)] + \min[Q(\phi)]},
	\label{eq:charge_visibility_definition}
\end{equation}
where we have to maximize and minimize the transferred charge over the phase difference $\phi$.
To evaluate the visibility from Eq.~(\ref{eq:charge_visibility_definition}), we again decompose the charge as $Q = Q_\text{cl} + Q_\text{int}$ and consider the two terms seperately. For $\Delta\tau \neq 0$, the interference term can be expressed as 
\begin{equation}
	Q_\text{int}(\phi) = -2e\sqrt{R_1R_2T_1T_2}\chi\left(\frac{\pi\Delta\tau}{\hbar\beta}\right)\left|G(\tau_u, \tau_d)\right| \cos[\alpha_2 - \alpha_1(\phi)],
	\label{eq:Q_int_phi}
\end{equation}
where we have introduced the phase $\alpha_1(\phi) = k_F\Delta L - \phi$, defined the function
 \begin{equation}
	G(\tau_u, \tau_d) \equiv \int G^{(1)}(t-\tau_u, t-\tau_d)dt,
	\label{eq:integrated_G1}
\end{equation} 
and then written $G(\tau_u, \tau_d)$ in polar form with the complex phase $\alpha_2$. 
As a result, the integral reduces to
\begin{equation}
		\int \text{Re}\left\{ \e^{-i\alpha_1(\phi) }G^{(1)}(t-\tau_u, t-\tau_d) \right\}dt= \left|G(\tau_u, \tau_d)\right| \cos[\alpha_2 - \alpha_1(\phi)].
	\label{eq:G_integral}
\end{equation}
From Eq.~(\ref{eq:Q_int_phi}), we see that~$Q_\text{int}$ only depends on the phase $\phi$ through the cosine term, which takes values between -1 and 1. Thus, for the maximum value of $Q_\text{int}$, we get       
\begin{equation}
	\text{max}(Q_\text{int}) = 
	2e\sqrt{R_1R_2T_1T_2} \chi\left(\frac{\pi\Delta\tau}{\hbar\beta}\right) \left| G(\tau_u, \tau_d) \right|,  \quad \Delta\tau \neq 0,
	\label{eq:max_transmitted_charge_1}
\end{equation}
while for the minimum, we have
\begin{equation}
	\text{min}(Q_\text{int}) = -\text{max}(Q_\text{int}).
	\label{eq:q_max_min}
\end{equation} 
For $\Delta\tau = 0$, we have 
\begin{equation}
	Q_\text{int}(\phi) = -\frac{2e^2}{h}\sqrt{R_1R_2T_1T_2}\cos(\phi) \int V(t) dt,
\end{equation}
and therefore 
\begin{equation}
	\text{max}(Q_\text{int}) = 
	\frac{2e^2}{h}\sqrt{R_1R_2T_1T_2}\left|\int V(t) dt\right|, \quad  \Delta\tau = 0
\end{equation}
together with Eq.~(\ref{eq:q_max_min}). Moreover, since 
$Q_\text{cl}$ is independent of $\phi$, we immediately get $\text{max}(Q_\text{cl}) = \text{min}(Q_\text{cl}) = Q_\text{cl}$, and combining these results, we obtain the expression
\begin{equation}
	\nu = \frac{\text{max}(Q_\text{int})}{Q_\text{cl}} = 
	\begin{cases}
		\nu_0 \frac{h}{e}\frac{\left| G(\tau_u, \tau_d) \right|}{\int V(t) dt}, & \Delta\tau \neq 0\\
		\nu_0 \frac{\left|\int V(t) dt\right|}{\int V(t) dt}, &  \Delta\tau = 0\\ 
	\end{cases} ,
	\label{eq:charge_visibility_2}
\end{equation}
where we have defined the prefactor
\begin{equation}
	\nu_0  = 2\frac{\sqrt{R_1R_2T_1T_2}}{R_1R_2 + T_1T_2}\chi\left(\frac{\pi\Delta\tau}{\hbar\beta}\right),
\end{equation}
which is independent of the voltage pulses.

\section{Results \& analysis}
\label{sec:visi}

We are now ready to analyze the charge transferred through the Mach-Zehnder interferometer and the corresponding interference effects. We have already seen how the charge visibility is captured by the absolute value of the time-averaged excess correlation function, cf.~Eqs.~(\ref{eq:integrated_G1}) and (\ref{eq:charge_visibility_2}). 
We now go on to show how the properties of the correlation function, in particular the possibility to decompose it into elementary components, can explain the interference patterns in the visibility. 

\subsection{Fraunhofer-like diffraction pattern}

\begin{figure}
	\centering
	\includegraphics[width=1.0\textwidth]{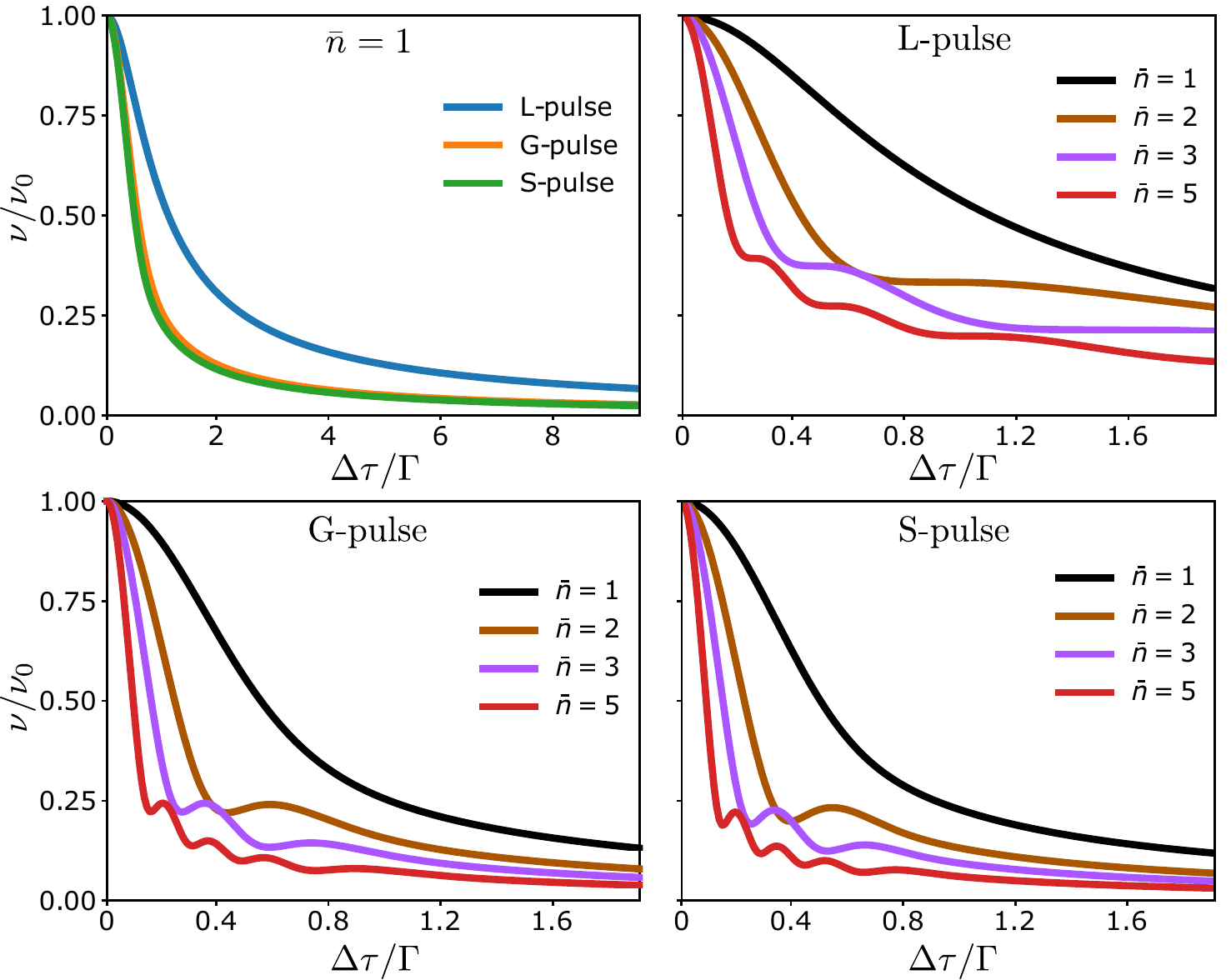}
	\caption{
		Visibility for different pulse types. Results are shown as a function of the difference of travel times $\Delta\tau$ over the pulse width~$\Gamma$ for Lorentzian (L), Gaussian (G), and sine (S) pulses with different number of charges~$\bar{n}=1,2,3,5$. 
	}
	\label{fig:single_pulse}
\end{figure}

We start by considering a single voltage pulse, and in Fig.~\ref{fig:single_pulse} we show the visibility given by Eq.~(\ref{eq:charge_visibility_2}) as a function of the difference of travel times $\Delta\tau$ over the pulse witdh $\Gamma$. We consider Lorentzian pulses, Eq.~(\ref{eq:lorentzian_voltage_pulse}),  Gaussian pulses,  Eq.~(\ref{eq:gaussian_voltage_pulse}), and sinusoidal pulses,  Eq.~(\ref{eq:sinusoidal_voltage_pulse}). In each case, we see that the visibility decreases as a function of the difference of travel times. 
If the difference of travel times is large compared to the width, the parts of the wave function going through different arms do not interfere when they reach the second quantum point contact, leading to a suppression of the visibility. 
By contrast, for short differences, of travel times compared to the pulse width, a pulse can interfere with itself at the second quantum point contact.

The visibility for Gaussian and sinusoidal pulses are nearly identical and qualitatively similar to the Lorentzian pulses. The difference between the Lorentzian and the other pulses is mainly due to the long tails of the Lorentzian ones, which extend beyond $|t|\geq\Gamma$. The similarity between the Gaussian and the sinusoidal pulses suggests that the exact shape of the pulses only has a small influence on the visibility. Owing to the similar behavior of the different pulses, we  focus now on the Lorentzian ones, which can be interpreted in terms of elementary excitations that leave the Fermi sea unaltered. 

Continuing with a single pulse, we now allow it to carry more than a single charge. The visibility is quite featureless for $\bar{n}=1$ according to Fig.~\ref{fig:single_pulse}. By contrast, more interesting structures appear as $\bar{n}$ is increased. For $\bar{n}>1$, oscillations appear in the visibility, forming a Fraunhofer-like interference pattern (a peak with superimposed interference oscillations in the tails) as seen in Fig.~\ref{fig:single_pulse}, with the number of oscillations being proportional to $\bar{n}$. To understand this Fraunhofer-like diffraction pattern, we analyze the correlation function $G^{(1)}$ in more detail. 

All voltages in Fig.~\ref{fig:single_pulse} are proportional to $\bar{n}$, and we can write the phase in Eq.~(\ref{eq:j_j_star}) as $\Phi(t_1, t_2) = -\bar{n}\varphi(t_1, t_2)$ with $\varphi$ being the phase for a single pulse with $\bar{n}=1$. Hence, for integer values of $\bar{n}$, we can decompose the numerator of the correlation function as
\begin{equation}
	\begin{split}
		\e^{-i\Phi(t_1, t_2)} - 1 &= (\e^{i\varphi(t_1, t_2)} - 1)(\e^{i(\bar{n}-1)\varphi(t_1, t_2)} + \e^{i(\bar{n}-2)\varphi(t_1, t_2)} + ... + 1) \\
		&= \sum_{k=0}^{\bar{n} - 1}(\e^{i\varphi(t_1, t_2)} - 1)\e^{ik\varphi(t_1, t_2)}, 
	\end{split}
	\label{eq:phi_derivation}
\end{equation}
When combined with Eq.~(\ref{eq:G1_envelope}),  we can then write the correlation function as
\begin{equation}
	G^{(1)}(t_1, t_2) = g(t_1, t_2)\sum_{k=0}^{\bar{n} - 1} \e^{ik\varphi(t_1, t_2)}
	\label{eq:G_single_pulse}
\end{equation}
with 
\begin{equation}
	g(t_1, t_2) = \frac{\e^{i\varphi(t_1, t_2)} - 1}{2\pi i(t_1 - t_2)}.
	\label{eq:g_def}
\end{equation}
We note that $g(t_1, t_2)$ is the correlation function of a single pulse with $\bar{n}=1$. Consequently, Eq.~(\ref{eq:G_single_pulse}) implies that the correlation function $G^{(1)}$ for a pulse with integer $\bar{n}>1$ is a sum of $\bar{n}$ elementary correlation functions, which differ from each other only by the relative phases $k\varphi$ with $k=0,\ldots,\bar{n} - 1$. 

The visibility is proportional to the time-integrated correlation function $G(\tau_u, \tau_d)$ according to Eq.~(\ref{eq:integrated_G1}). We thus denote the correlation function $G(\tau_u, \tau_d)$ for a single pulse with charge $\bar{n}$ as $G_{\bar{n}}(\tau_u, \tau_d)$. 
Using Eq.~(\ref{eq:G_single_pulse}) and omitting the dependence on $\tau_u$ and $\tau_d$ to keep the notation simple, we find
\begin{equation}
		G_{\bar{n}} = \int  dt g\sum_{k=0}^{\bar{n} - 1} \e^{ik\varphi }
		= \sum_{k=0}^{\bar{n} - 1}\mathcal{G}_{k}\e^{i\theta_k}
		= \mathcal{G}_0e^{i\theta_0}\sum_{k=0}^{\bar{n} - 1}\frac{\mathcal{G}_k}{\mathcal{G}_0}\e^{i(\theta_k-\theta_0)},
	\label{eq:integrated_G1_single_pulse}
\end{equation}
where we have defined the real numbers $\mathcal{G}_k$ and $\theta_k$ such that $\mathcal{G}_ke^{i\theta_k} = \int dt ge^{ik\varphi} $. In analogy with Eq.~(\ref{eq:G_single_pulse}), we can now interpret this expression as a sum of elementary contributions, where each term of the sum has a different phase $\theta_k-\theta_0$ and amplitude $\mathcal{G}_k/\mathcal{G}_0$. 
Hence, the interference effects can be understood as interference between the correlation functions of the elementary excitations within the pulse. We refer to this as interference of the second kind in contrast to conventional interference, where it is the wave function that interferes. We stress that while the wave function of a single particle can interfere with itself only, the correlation functions of various single particles can interfere with each other. 
Specifically, the Fraunhofer-like diffraction pattern in Fig.~\ref{fig:single_pulse} can be captured by the expression
\begin{equation}\label{eq:visibility_single-pulse}
	\frac{\nu}{\nu_0} = \frac{1}{\bar{n}} \left| \sum_{k=0}^{\bar{n} - 1} \mathcal{G}_k \e^{i(\theta_k-\theta_0)} \right| ,
\end{equation}
which is valid for $\Delta\tau\neq0$ and $\bar{n}$ being an integer. 

\begin{figure}
	\centering
	\includegraphics[width=0.95\textwidth]{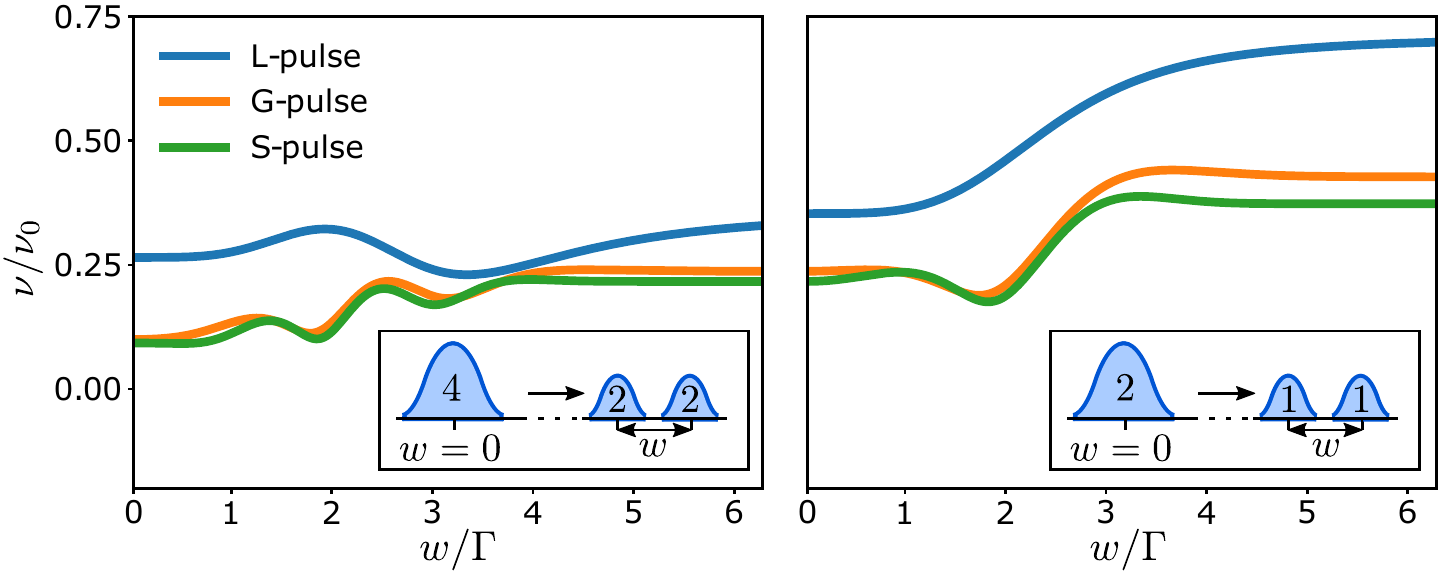}
	\caption{Visibility for two charge pulses with $\bar{n} = 2$ (left panel) and $\bar{n} = 1$ (right panel). The time delay between the charge pulses is denoted by $w$, and we have used $\Delta\tau / \Gamma=0.6$. }
	\label{fig:two_pulse}
\end{figure}

In addition to the interference oscillations, we see that the overall decay of the visibility in Fig.~\ref{fig:single_pulse} for $\bar{n}>1$  is similar to the case $\bar{n}=1$. Generally, the visibility decreases as a function of the difference of travel times, but the decay is faster for larger values of $\bar{n}$. Thus, the interference caused by the relative phases tend to reduce the visibility. From Fig.~\ref{fig:single_pulse} we also conclude that to observe interference, the difference of travel times, $\Delta\tau$, and the width, $2\Gamma$, should be of the same order.  For $\Delta \tau \ll \Gamma$ or $\Delta \tau \gg \Gamma$, there are no oscillations in the visibility. 

\subsection{Diffraction grid}

Having established that the interference of the elementary components in a single pulse is revealed in the visibility as a Fraunhofer-like diffraction pattern, we now consider two separated pulses with arbitrary charge. In the case with only a single pulse, the interference pattern was fully determined by the charge of the pulse and the difference of the arm lengths. Now, the time delay $w$ between the two pulses, see Eq.~(\ref{eq:finite-train}), offers another way to control the interference pattern. Figure~\ref{fig:two_pulse} shows the effect of changing the delay $w$ between two pulses with charge $\bar{n}=2$ ($\bar{n}=1$) in the left (right) panel. For $w=0$, the pulses combine into a single pulse with the charge given by the total charge of the pulses. As the time delay in increased, the pulse splits into the two separate pulses. As we further decompose the multi-charged pulse into its elementary parts with $\bar{n}=1$, the number of oscillations decreases and the visibility is enhanced as can be seen by comparing the left and right panels. We also note that if we take $w \gg \Gamma$, we completely split the pulses and they do not interfere. In Fig.~\ref{fig:two_pulse}, the oscillations are concentrated near $w=\Delta\tau=2\Gamma$, and we should generally choose $w$ close to those values to observe the clearest interference effects. 

\begin{figure}
	\centering
	\includegraphics[width=0.7\textwidth]{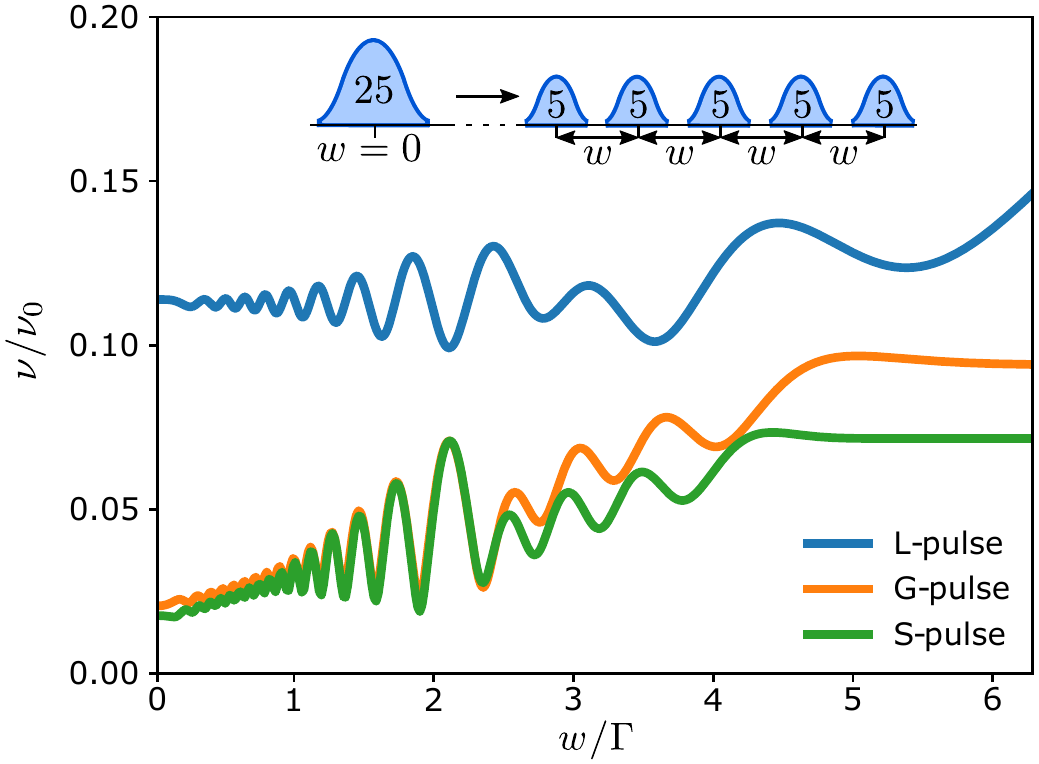}
	\caption{Visibility for five voltages pulses of different types with $\bar{n} = 5$ particles each. The time delay between the pulses is denoted by $w$, and we have used $\Delta\tau / \Gamma=0.6$.
	}
	\label{fig:five_pulses}
\end{figure}

Under these conditions, we can further enhance the interference effects by increasing both the number of pulses $m$ and their charge $\bar{n}$. This leads to an increased number of oscillations as seen in Fig.~\ref{fig:five_pulses}, where we consider the splitting of a single pulse with $\bar{n}=25$ into five pulses with $\bar{n}=5$. To maximize the Fraunhofer-like pattern, the pulses are separated by the same time delay $w$. The magnitude of the visibility and its oscillations are reduced by the increased interference.  
Again, the visibility of a single pulse is given by its width $\Gamma$ and its total charge $\bar{n}$. Figure \ref{fig:five_pulses} shows how the visibility starts to rapidly oscillate with the time delay $w$ as we decompose the large pulse into smaller ones. This behavior resembles the diffraction pattern generated by a regular spatial grid, where the smaller pulses here are acting like the slits of the grid, which are spaced by the delay time $w$. Thus, by tuning the separation between the pulses, one can directly control the resulting diffraction pattern.

\section{Conclusions}
\label{sec:conc}

We have theoretically investigated multi-particle interference  in an electronic Mach-Zehnder interferometer driven by dynamic voltage pulses. To this end, we have described a Floquet scattering theory, which allows us to calculate the time-dependent current and the excess correlation function in a mesoscopic conductor driven by a time-dependent voltage. The current can be divided into a classical part and an interference term, with the classical contribution being independent of the temperature, while the interference term  is gradually washed out by an increasing temperature. Moreover, the interference term determines the visibility that we have
investigated for different pulse types.

For a single Lorentzian-shaped voltage pulse containing several charges, we have shown how oscillations in the visibility can be related to interferences between the individual charges making up the pulse. These interferences give rise to a Fraunhofer-like diffraction pattern consisting of interference oscillations superimposed on a central peak structure. Additional features appear as several pulses are injected into the interferometer, making it possible for different pulses to arrive simultaneously at the second quantum point contact of the setup. In that case, we observe an interference pattern that resembles what one would get with a regular spatial grid, however, with the spacing between the slits of the grid replaced by the delay time between the pulses. These predictions may be observed in future experiments by injecting multi-particle pulses into an electronic Mach-Zehnder inteferometer.

\vspace{6pt} 

\authorcontributions{J.~K., P.~B.,~and M.~M.~performed calculations, and C.~F.~supervised the research. All authors discussed and analyzed the results and contributed to the writing of the manuscript.}


\acknowledgments{We thank P.~Portugal and B.~Roussel for useful discussions. M.~M.~acknowledges the hospitality of Aalto University and support from Aalto Science Institute through its Visiting Fellow Programme. We acknowledge the computational resources provided by the Aalto Science-IT project. }


\appendixtitles{no} 
\appendixsections{multiple} 

\end{document}